\documentclass[12pt]{article}

\usepackage[top=1in,bottom=1in,left=1.25in,right=1.25in]{geometry}

\usepackage[T1]{fontenc}
\usepackage[utf8]{inputenc}

\usepackage{amsmath,amssymb}

\usepackage{booktabs}
\usepackage{multirow}
\usepackage{array}
\usepackage{threeparttable}   
\newcolumntype{L}[1]{>{\raggedright\arraybackslash}p{#1}}
\newcolumntype{C}[1]{>{\centering\arraybackslash}p{#1}}
\newcolumntype{R}[1]{>{\raggedleft\arraybackslash}p{#1}}
\usepackage{float}
\usepackage{graphicx}
\graphicspath{{./}{figures/}}

\usepackage{setspace}
\usepackage{fancyhdr}
\usepackage{titlesec}
\titleformat{\section}{\large\bfseries}{\thesection.}{0.5em}{}
\titleformat{\subsection}{\normalsize\bfseries}{\thesubsection}{0.5em}{}
\titlespacing*{\section}{0pt}{18pt}{6pt}
\titlespacing*{\subsection}{0pt}{12pt}{4pt}

\usepackage[font=small,labelfont=bf,
            justification=justified,singlelinecheck=false]{caption}

\usepackage[colorlinks=true,
            linkcolor=black,
            citecolor=black,
            urlcolor=black]{hyperref}

\usepackage{natbib}

\usepackage[bottom]{footmisc}

\usepackage{xcolor}
\usepackage{enumitem}
\begin{document}

\begin{titlepage}
\begin{center}
\vspace*{1.5cm}
{\LARGE\bfseries Macro Economists in the Machine:\\[6pt]
A Multi-Agent LLM Framework for\\[4pt]
Commodity-Related ETF Portfolio Construction}\\[2cm]

{\large
Yiqing Wang\textsuperscript{a} \quad
Dehao Dai\textsuperscript{b} \quad
Ding Ma\textsuperscript{c} \quad
Kerui Geng\textsuperscript{d,*}
}\\[1cm]

{\normalsize
\textsuperscript{a}Citigroup, Dallas, Texas\\[3pt]
\textsuperscript{b}University of California San Diego, La Jolla, California\\[3pt]
\textsuperscript{c}Georgia Institute of Technology, Atlanta, Georgia\\[3pt]
\textsuperscript{d}Tulane University, New Orleans, Louisiana\\[8pt]
\textsuperscript{*}Corresponding author: \href{mailto:kerui.geng@tulane.edu}{kgeng@tulane.edu}
}\\[1.5cm]

{\normalsize This version: \today}

\end{center}
\end{titlepage}

\newpage
\thispagestyle{plain}
\begin{center}
{\large\bfseries Abstract}
\end{center}
\vspace{0.5cm}

\noindent
We test whether large language models (LLMs) add value in commodity portfolio construction when the information set and implementation rules are held fixed across strategies.
A Hawkish Agent (inflation-tightening prior), a Dovish Agent (growth-easing prior), a Debate Agent, and a deterministic z-score Rule Agent each receive identical FRED macro z-scores and route their tilt signals through the same portfolio engine.
Across 124 weekly rebalancing dates spanning the 2023 U.S.\ rate peak and the 2024--2025 soft landing, all three LLM strategies outperform the Rule Agent in Sharpe terms; the Hawkish and Debate Agents record the largest gains ($\Delta\text{Sharpe} = +0.044$ and $+0.040$, both $p < 0.10$ under a block bootstrap) and preserve a net-of-cost advantage over the passive inverse-volatility benchmark at one-way trading costs up to 30 basis points, while the Rule Agent's thin margin over passive disappears at approximately 5 basis points.
The Debate Agent does not outperform the best single agent ($\Delta\text{Sharpe} = -0.004$, $p = 0.769$); its contribution is bias correction---averaging out the Dovish Agent's miscalibrated prior---rather than deliberation-generated return.
The performance advantage is concentrated in the soft-landing sub-period, the evaluation window spans a single rate cycle, and the reported $p$-values are unadjusted for multiple comparisons.
Within these limits, the results suggest that an LLM acting as a constrained macro-interpretation function can add modest but economically meaningful value over a transparent rule layer, though the margin is small and its persistence beyond this sample is unknown.

\vspace{0.8cm}
\newpage


\section{Introduction}
\label{sec:intro}
Commodity allocation as a traditional asset allocation is an investment strategy in which investors construct portfolios of commodity exposures to maximize risk-adjusted returns and to hedge against inflation and macroeconomic shocks. In recent years, investors and trading institutions have increasingly adopted quantitative strategies to make decisions. This shift has been driven by the growing availability of high-frequency price data, the maturation of liquid commodity ETFs and futures markets, and a substantial empirical literature linking commodity returns to identifiable risk premia.

A first generation of quantitative strategies, rooted in factor investing, exploits the cross-sectional and time-series structure of commodity returns. Momentum, carry, basis, hedging pressure, and value have been documented as robust drivers of commodity futures returns \citep{erb2006strategic, gorton2006facts, gorton2013fundamentals,daskalaki2014there, szymanowska2014anatomy, bakshi2019understanding, dai2025statistical}. These factors are constructed from ranked signals and provide a systematic, replicable structure for commodity portfolio construction. More recently, machine learning techniques have emerged as alternatives to classical factor models, with tree-based ensembles, neural networks, and deep learning architectures applied to commodity return forecasting and dynamic portfolio construction \citep{gu2020empirical, jiang2017deep, kelly2023financial}. Compared to the classical factor models, these methods can relax the linearity and stationarity assumptions to capture nonlinear interactions and high-dimensional dependencies between predictors. However, commodity markets exhibit pronounced non-stationarity, structural breaks driven by geopolitics and the energy transition, and heavy-tailed returns, all of which limit the in-sample patterns these methods can reliably learn.

Large language models (LLMs) are now widely used to process the unstructured information flow related to commodity prices, such as central bank communications, OPEC press releases, or geopolitical news. LLMs can extract sentiment signals from textual data \citep{ kim2024financial, dai2026beyond}, synthesize heterogeneous evidence into structured outputs such as risk indices or scenario forecasts \citep{ yang2023fingpt, lopez2023can}, and make decisions within agentic trading frameworks\citep{wu2023bloomberggpt}. Multi-agent designs, in which several LLM agents interact through structured deliberation, have attracted particular attention as a way to combine specialization with disciplined aggregation \citep{chan2023chateval, liang2024encouraging}. However, LLMs in these applications often appear useful in part because they have access to richer text inputs and more flexible trading systems than the rule-based baselines they are compared with, making it difficult to isolate the contribution of the language model itself. A clean methodological question arises: given the same structured macro data and the same portfolio engine, can a large language model produce a better mapping from macro conditions to commodity ETF tilts than a deterministic rule?

Commodity allocation is a particularly demanding setting for this test since its macroeconomic drivers often pull in opposite directions, and a fixed sign-mapping cannot resolve such conflicts. The empirical design keeps the comparison clean. At each weekly rebalancing date, all strategies observe the same seven indicators: the VIX, the broad U.S. dollar index, the federal funds rate, industrial production, 10-year breakeven inflation, the 10-year TIPS real yield, and the unemployment rate. Each variable is converted into a rolling three-year $z$-score. The strategies differ only in how they translate these $z$-scores into ticker-level tilts; after that step, every strategy uses the same inverse-volatility base allocation, equal-weight blend, macro-tilt scaling, risk controls, and turnover constraints. The exercise is therefore an ablation of the interpretation layer, not a comparison of unrestricted trading systems.

We compare four signal-generating approaches. The Rule Agent applies a pre-specified loading matrix that maps each macro $z$-score to fixed tilts on the underlying ETFs. The Hawkish Agent places greater weight on inflation control, tight monetary policy, and elevated real rates, while the Dovish Agent places greater weight on employment, growth support, and recovery momentum. The Debate Agent combines the two single-agent outputs through a structured review step. This design varies the macro prior across agents while holding the rest of the implementation fixed, and it allows us to test whether the debate structure adds value beyond either single prior. All three LLM agents outperform the Rule Agent in full-sample Sharpe terms, with the largest gains for the Hawkish and Debate Agents. The bootstrap evidence for these two comparisons reaches the 10\% level, although the confidence intervals include zero and the gains do not survive a conservative multiple-testing correction. The advantage is also regime-dependent: it is concentrated in the 2024--2025 soft-landing period, when inflation moderated while growth remained resilient, whereas during the 2023 rates-peak period the passive inverse-volatility benchmark outperforms all signal-based strategies. Within the LLM family, the Debate Agent does not outperform the stronger single agent, suggesting that its main role is to hedge against a poorly aligned prior rather than to generate a separate deliberation premium.

Transaction-cost tests support the economic relevance of these findings within reasonable cost levels. The Hawkish and Debate Agents retain a Sharpe advantage over the inverse-volatility benchmark at one-way costs up to 30 basis points, while the Rule Agent's small advantage over the same benchmark largely disappears even at low cost levels. The performance differences are nonetheless small, and the sample is short. The paper should therefore be read as evidence that LLM-based macro interpretation, embedded in a controlled portfolio engine, can improve on a rule-based mapping in this setting, not as evidence of persistent trading skill.

The paper makes three contributions. First, it extends work on LLMs in finance \citep{lopez2023can, wu2023bloomberggpt, kim2024financial} from text-based return prediction to constrained portfolio construction with structured macro inputs. Second, it adds to the multi-agent LLM literature \citep{chan2023chateval, liang2024encouraging} by treating Hawkish and Dovish macro priors as testable design choices rather than free stylistic variation. Third, it speaks to the commodity allocation literature by asking whether LLM-based macro interpretation improves on rule-based baselines built from familiar macro predictors. The closest related work is \citet{zhao2025alphaagents}; our design is narrower because it isolates the macro-interpretation layer in a setting where the macroeconomic channels driving commodity returns are direct and well documented.

The remainder of the paper is organized as follows.
Section~\ref{sec:literature} reviews related work on LLMs in finance, multi-agent designs, and commodity allocation. Section~\ref{sec:methodology} describes the agent architecture and the portfolio construction procedure that all four strategies share. Section~\ref{sec:data} presents the macro indicators and ETF universe used in the empirical exercise. Section~\ref{sec:results} reports the main performance comparisons, subperiod analysis, and transaction-cost robustness. Section~\ref{sec:discussion} discusses the findings and their limitations. Section~\ref{sec:conclusion} summarizes the comparison across agents and discusses implications for LLM-based allocation strategies.

\section{Related Literature}
\label{sec:literature}

\subsection{Commodity Allocation and Macroeconomic Drivers}

Commodity allocation has long been studied as both a strategic asset allocation problem and a tactical source of macroeconomic exposure.Early studies document that commodity futures can provide diversification benefits relative to traditional equity and bond portfolios, while also exhibiting distinct risk-return behavior across macroeconomic regimes \citep{gorton2006facts, erb2006strategic}. 

A related strand of work identifies the cross-sectional and time-series drivers of commodity returns. Term structure, basis, hedging pressure, momentum, carry, and inventory-related variables have been shown to explain meaningful variation in commodity futures returns \citep{gorton2013fundamentals, szymanowska2014anatomy, daskalaki2014there, bakshi2019understanding}. These findings motivate rule-based and factor-based commodity strategies, in which transparent signals are mapped into portfolio weights using economically interpretable relationships. For example, backwardation and positive carry are commonly associated with higher expected commodity returns, while contango and weak demand conditions are often interpreted as unfavorable signals. Such approaches are attractive because they are transparent, auditable, and easy to implement.

Macroeconomic variables provide another important channel for commodity allocation. Inflation expectations, real interest rates, the U.S. dollar, industrial production, financial-market volatility, and labor-market conditions all affect commodity-related assets through different mechanisms. Higher inflation expectations may support precious metals and broad commodity exposures, while higher real yields increase the opportunity cost of holding non-yielding assets such as gold. A stronger U.S. dollar can pressure dollar-denominated commodities, whereas stronger industrial production may support energy, metals, and other cyclical commodities. Similarly, elevated market volatility can signal risk-off conditions that reduce the attractiveness of cyclical commodity exposures. These channels motivate the use of macro variables such as the VIX, the broad dollar index, the federal funds rate, industrial production, breakeven inflation, real yields, and unemployment as state variables for commodity-related allocation decisions.

However, a key difficulty is that commodity-relevant macro signals often point in different directions. For example, an environment with elevated inflation expectations may appear favorable for commodities, but if it is accompanied by high real yields, a strong dollar, or weak industrial activity, the directional implication becomes ambiguous. A deterministic sign-mapping rule can encode standard macro-finance relationships, but it may be too rigid when signals conflict. This limitation is especially relevant after the financialization of commodity markets, as index participation and cross-asset flows have increased the interaction between commodity returns, equity markets, and macro-financial conditions \citep{tang2012index}. As a result, adaptive interpretation of macro states may be valuable even when the underlying macro variables are familiar and economically motivated.

Our paper builds on this literature but asks a different question. We do not test whether macro variables have predictive power for commodity returns in general. Instead, we examine whether a constrained large language model can translate the same standardized macro state into better commodity-related ETF tilts than a fixed rule-based mapping. This distinction is important because the information set, asset universe, portfolio construction engine, and risk controls are held constant across strategies. The comparison therefore isolates the incremental value of the macro-interpretation layer rather than the value of a different forecasting model or a more flexible portfolio optimization system.

\subsection{Machine Learning and LLMs in Financial Signal Generation}

A second related literature studies the use of machine learning and artificial intelligence for financial signal generation and asset allocation. Traditional rule-based strategies are attractive because they are transparent, easy to audit, and closely tied to economic intuition. However, fixed rules typically impose stable directional relationships between predictors and assets. This can be restrictive in financial markets, where predictive relationships are often nonlinear, state-dependent, and unstable across regimes. Machine learning methods have therefore been introduced to capture complex interactions among predictors and to improve return prediction and portfolio construction. For example, \citet{gu2020empirical} shows that machine learning methods can improve empirical asset-pricing performance by capturing nonlinear relationships in a high-dimensional predictor space. \citet{kelly2023financial} provide a broad overview of financial machine learning and emphasize its ability to model flexible predictive structures, while also noting the challenges created by limited samples, non-stationarity, and overfitting. Deep learning and reinforcement learning methods have also been applied to dynamic portfolio allocation, where models learn allocation policies directly from market data \citep{jiang2017deep}.

Large language models provide a distinct approach to financial signal generation. Instead of learning only from numerical return histories, LLMs can process language, summarize heterogeneous information, and translate qualitative evidence into structured signals. This development builds on an earlier textual-analysis literature showing that language contains economically meaningful information. \citet{tetlock2007} shows that media pessimism predicts stock-market movements and trading volume, while \citet{loughran_mcdonald_2011} demonstrate that financial-domain dictionaries are necessary because generic sentiment classifications often misrepresent the meaning of words in corporate disclosures. More recent work studies whether GPT-class models can extract return-relevant information from financial text. \citet{lopez2023can} find that ChatGPT-based sentiment scores on news headlines contain predictive information for short-horizon stock returns. \citet{kim2024financial} show that large language models can perform financial statement analysis tasks and generate useful forecasts from structured accounting information. Domain-specific financial LLMs, such as BloombergGPT and FinGPT, further illustrate the growing role of LLMs in financial natural language processing and decision support \citep{wu2023bloomberggpt, yang2023fingpt}.

Most LLM-finance studies focus on information extraction from unstructured text, such as news, earnings calls, filings, or analyst reports. In these settings, the LLM may add value because it has access to richer information or because it processes text more effectively than traditional sentiment models. Our setting is intentionally different. The LLM agents do not read unrestricted news, search for external information, or forecast asset returns directly. Instead, they receive a compact table of standardized macroeconomic indicators and produce bounded ticker-level tilt signals. The model's role is therefore limited to translating a structured macro state into portfolio tilts.

This distinction is important for identification. If an LLM-based trading strategy outperforms a rule-based benchmark while using richer inputs, different risk controls, or a more flexible trading engine, it is difficult to determine whether the improvement comes from the language model itself or from the broader system design. In our framework, the information set, asset universe, inverse-volatility base allocation, risk constraints, and transaction-cost assumptions are held fixed across strategies. The comparison therefore isolates whether LLM-based interpretation of the same macro evidence improves on a deterministic rule-based mapping. In this sense, the LLM is not treated as an unrestricted trading engine, but as a constrained interpretation layer embedded within a transparent portfolio construction framework.

\subsection{Multi-Agent LLM Systems and Financial Agents}

The intellectual motivation is related to the view that complex intelligence may arise from interactions among specialized sub-agents rather than from a single centralized mechanism \citep{minsky_1986}. In the LLM context, multi-agent debate has been proposed as a way to improve reasoning, factuality, and robustness. \citet{du_etal_2023} show that multiple language-model instances can propose, debate, and revise answers over several rounds, improving factual accuracy and mathematical reasoning relative to single-pass generation. \citet{liang2024encouraging} argue that debate among agents can encourage divergent thinking and reduce the tendency of a single model to remain locked into an initially incorrect reasoning path. Relatedly, \citet{chan2023chateval} introduce ChatEval, a multi-agent debate framework for LLM-based evaluation, and show that multiple agent-judges can improve open-ended text evaluation relative to single-agent judging.

Recent work has extended the multi-agent paradigm to financial decision-making. \citet{zhao2025alphaagents} propose AlphaAgents, a role-based multi-agent framework for equity portfolio construction in which specialized agents collaborate on stock selection and portfolio formation. \citet{xiao_etal_2024} introduce TradingAgents, a multi-agent trading framework that simulates the structure of a trading firm through agents such as fundamental analysts, sentiment analysts, technical analysts, bull and bear researchers, traders, and risk managers. These studies highlight the potential of agent specialization and structured debate in investment workflows.

However, existing financial multi-agent systems are often broad trading architectures rather than controlled tests of the value of multi-agent reasoning itself. Agents may process heterogeneous data sources, search for external information, use domain-specific tools, or influence several stages of the trading process. As a result, superior performance may reflect richer information, more flexible portfolio construction, or stronger risk management rather than the incremental value of multi-agent interpretation.

Our framework is intentionally more restricted. The Hawkish and Dovish agents observe the same standardized macroeconomic state and differ only in their interpretive prior. The Hawkish Agent places greater weight on inflation control, tight monetary policy, and elevated real rates, while the Dovish Agent places greater weight on employment, growth support, and recovery momentum. Neither agent can search for additional information, call external tools, or modify the downstream portfolio construction engine. The Debate Agent then aggregates the two single-agent outputs through a structured review mechanism. This design allows us to study multi-agent debate as a controlled prior-aggregation mechanism rather than as an unrestricted trading architecture.

This distinction is important for interpreting the empirical results. If the Debate Agent outperforms the Rule Agent, the improvement can be attributed to the mapping from the same macro state into ETF-level tilt signals, not to a different information set or a different portfolio engine. Conversely, if the Debate Agent does not outperform the stronger single agent, the result suggests that debate mainly stabilizes exposure across competing macro priors rather than creating an independent deliberation premium. Thus, our contribution to the multi-agent LLM literature is to evaluate whether structured disagreement between macroeconomic priors adds value in a constrained portfolio-construction setting.

\subsection{Controlled Portfolio Evaluation and the Interpretation Layer}

A central challenge in evaluating AI-based trading systems is attribution. When an AI strategy outperforms a benchmark, the source of the improvement is often unclear. It may reflect a richer information set, a more flexible forecasting model, a different portfolio optimizer, stronger risk controls, or a favorable sample period rather than the specific contribution of the AI component itself. This problem is especially important in portfolio applications, where small changes in asset universe, weighting rule, turnover constraint, or transaction-cost assumption can materially change measured performance.

The empirical asset-pricing and portfolio-choice literature has long emphasized the importance of transparent benchmarks and disciplined out-of-sample evaluation. Simple allocation rules, such as equal-weighted or inverse-volatility portfolios, are commonly used because they provide robust reference points and avoid excessive estimation error. For example, \citet{demiguel_garlappi_uppal_2009} show that naive diversification can be difficult to outperform out of sample, while \citet{demiguel_garlappi_nogales_uppal_2009} emphasize that portfolio constraints can improve performance by limiting estimation error and extreme portfolio weights. These findings motivate the use of simple, transparent portfolio engines as benchmarks when evaluating more flexible signal-generation methods.

A related literature cautions that financial backtests are vulnerable to data snooping, multiple testing, and overfitting. \citet{white_2000} develops a reality-check framework for evaluating whether a selected strategy truly outperforms a benchmark after accounting for data snooping. \citet{hansen_2005} proposes the superior predictive ability test as a refinement for comparing many forecasting models. \citet{bailey_etal_2014} emphasize that repeated strategy search can create backtest overfitting, causing apparently strong performance to disappear out of sample. In the asset-pricing context, \citet{harvey_liu_zhu_2016} similarly argue that multiple testing requires more conservative evidence thresholds when many candidate predictors or strategies are examined.

Our design responds to these concerns by separating the interpretation layer from the portfolio construction layer. The Rule Agent, Hawkish Agent, Dovish Agent, and Debate Agent all observe the same standardized macroeconomic variables and trade the same commodity-related ETF universe. After each strategy produces ticker-level tilt signals, the signals are passed through the same inverse-volatility base allocation, equal-weight blend, macro-tilt scaling rule, risk-off cap, single-name weight cap, turnover constraint, and transaction-cost calculation. The passive inverse-volatility portfolio further serves as a no-macro-signal benchmark. Therefore, differences in performance across the active strategies are not driven by different risk engines or different implementation rules.

This controlled design makes the empirical comparison narrower but more interpretable. The paper does not ask whether an unrestricted LLM trading agent can generate profitable commodity trades. Instead, it asks whether an LLM can improve the mapping from a fixed macroeconomic state to bounded ETF-level tilt signals relative to a deterministic rule. In this sense, the LLM is evaluated as a constrained interpretation layer embedded within a transparent portfolio framework. If the LLM agents outperform the Rule Agent, the improvement can be attributed to more flexible macro-state interpretation rather than to a larger information set or a different portfolio optimizer.

\subsection{Performance Inference in Short Financial Samples}

A final related literature concerns statistical inference for trading-strategy performance. Sharpe-ratio comparisons are widely used in empirical asset management, but their sampling properties are nontrivial in short samples and in the presence of serial dependence. \citet{lo_2002} shows that the Sharpe ratio has different asymptotic behavior under different return-generating processes and that the common square-root-of-time annualization rule can be misleading when returns are serially correlated. This issue is particularly relevant for portfolio strategies formed from overlapping macroeconomic conditions, rolling volatility estimates, turnover constraints, and weekly rebalancing rules.

Testing whether one strategy has a higher Sharpe ratio than another also requires care. Standard tests based on independent and normally distributed returns may be unreliable when returns are heavy-tailed or serially dependent. \citet{ledoit_wolf_2008} propose a robust bootstrap-based procedure for testing differences in Sharpe ratios and recommend time-series resampling methods to account for dependence in portfolio returns. This motivates our use of a paired stationary block bootstrap applied to the joint return vector across strategies. By resampling strategies jointly, the procedure preserves both the time-series dependence within each strategy and the contemporaneous dependence across strategies.

The stationary bootstrap provides a natural resampling method for weakly dependent financial time series because it resamples blocks of random length and preserves local serial dependence \citep{politis_romano_1994}. The choice of block length is important in finite samples, and automatic block-length selection procedures provide a data-dependent way to choose this tuning parameter \citep{politis_white_2004}. In our setting, this approach is useful because the evaluation window contains only 124 weekly observations, making asymptotic approximations potentially fragile.

A related concern is multiple testing. Portfolio studies often compare several strategies against multiple benchmarks, which can overstate evidence if unadjusted p-values are interpreted as confirmatory results. Stepdown multiple-testing methods such as those proposed by \citet{romano_wolf_2005_datasnooping} and \citet{romano2005exact} provide a formal way to control familywise error rates while accounting for dependence across test statistics. Because our sample is short and several pairwise Sharpe-ratio comparisons are reported, we interpret unadjusted bootstrap p-values as directional evidence rather than conclusive proof of persistent outperformance.

This inference framework is consistent with the paper's broader empirical objective. The goal is not to claim that LLM-based agents generate statistically robust trading profits in general. Instead, the bootstrap tests provide a disciplined way to assess whether the observed Sharpe-ratio differences are consistent with incremental value from the macro-interpretation layer, while recognizing the limitations created by short samples, serial dependence, and multiple comparisons.


\section{Methodology}
\label{sec:methodology}

\subsection{Architecture Overview}

This paper mainly compares the five strategies to construct an ETF portfolio. To ensure consistency and comparability across results, all strategies share the same data pipeline and the same portfolio construction engine. The following are the five strategies:  
\begin{enumerate}[leftmargin=*,label=(\arabic*)]
\item \textbf{Rule Agent}: a deterministic z-score mapping that serves as the transparent benchmark;
\item \textbf{Hawkish Agent}: an LLM agent with an inflation-control and monetary-tightening prior;
\item \textbf{Dovish Agent}: an LLM agent with a growth-supportive and monetary-easing prior;
\item \textbf{Debate Agent}: a consensus strategy formed from the Hawkish and Dovish outputs; and
\item \textbf{Inverse Volatility}: a passive risk-only benchmark with no macro signal.
\end{enumerate}

The important feature of the design is that the LLM agents do not receive additional information and do not control the final risk management step.
Their only role is to produce a macro tilt contract from the same set of inputs.
This restriction makes the comparison more conservative than a general AI trading exercise, but it also makes the source of any performance difference easier to interpret.

\subsection{Macro Feature Extraction}

At each weekly rebalancing date $t$, the system constructs a seven-dimensional feature vector from FRED:
\begin{equation}
\mathbf{f}_t = \bigl(\,z^{\text{vix}}_t,\; z^{\text{usd}}_t,\; z^{\text{ff}}_t,\; z^{\text{indpro}}_t,\; z^{\text{bkevn}}_t,\; z^{\text{ryr}}_t,\; z^{\text{unrate}}_t\,\bigr),
\label{eq:features}
\end{equation}
where each element is the rolling 156-week z-score of the corresponding macro series.
Standardizing the inputs serves two purposes.
It places variables with different units on a common scale, and it gives the LLM a simple description of whether each macro indicator is unusually high, unusually low, or close to normal.

The timing of the variables is designed to mitigate look-ahead bias. Daily market-based series, including the VIX, broad dollar index, breakeven inflation, and real yields, are measured using the latest observations available at or before each weekly rebalancing date. In contrast, lower-frequency macroeconomic releases, such as industrial production and unemployment, are incorporated only after their public release dates, so that each weekly feature vector reflects the information set available to investors at that time. Release-lag adjustments are applied before the rolling z-scores are computed. This procedure is release-aware rather than fully vintage-based: it does not reconstruct every historical data vintage or eliminate the effect of subsequent data revisions. Nevertheless, it avoids the most direct form of look-ahead bias by preventing unreleased macroeconomic data from entering the backtest.

\subsection{Rule Agent}
\label{sec:rule}
The Rule Agent translates the macro feature vector into portfolio tilt signals using a pre-specified loading matrix. For asset $i$, the raw tilt score is
\begin{equation}
\tilde{s}_{i,t}
=
\sum_{k \in \mathcal{K}}
\beta_{ik} z_{k,t}
\mathbf{1}\left\{ |z_{k,t}| \geq 0.5 \right\},
\label{eq:rule_raw_tilt}
\end{equation}
where $\beta_{ik} \in \{-1,0,+1\}$ denotes the directional exposure of asset $i$ to macro variable $k$. Macro deviations smaller than one-half standard deviation in absolute value are set to zero before aggregation, which reduces the influence of small and potentially noisy fluctuations. The raw score is then mapped into a five-level tilt signal,
\[
b_{i,t} \in \{-2,-1,0,+1,+2\},
\]
using thresholds at $\pm 1$ and $\pm 2$:
\begin{equation}
b_{i,t}
=
\begin{cases}
+2, & \tilde{s}_{i,t} \geq 2, \\
+1, & 1 \leq \tilde{s}_{i,t} < 2, \\
0, & |\tilde{s}_{i,t}| < 1, \\
-1, & -2 < \tilde{s}_{i,t} \leq -1, \\
-2, & \tilde{s}_{i,t} \leq -2.
\end{cases}
\label{eq:rule_discretization}
\end{equation}

For notational consistency with the portfolio construction step, the Rule Agent's final tilt signal is defined as
\[
s^{\mathrm{Rule}}_{i,t} = b_{i,t}.
\]
The loading matrix is reported in Appendix A. All loadings and thresholds are fixed ex ante and are not estimated from evaluation-period returns.

\subsection{LLM Agent Protocol}

Each LLM agent receives a structured prompt containing three elements: the agent's macro prior, the current table of standardized macro indicators, and a short machine-generated narrative summarizing the macro state.
The narrative is mechanically generated from the same standardized indicators and contains no external news, future returns, or future macro realizations. The agent returns a JSON contract with four components:
\begin{itemize}[leftmargin=*]
\item a probability distribution over macro regimes;
\item ticker-level tilt signals in $[-2,+2]$ for the 15 ETFs;
\item a brief economic rationale; and
\item confidence scores attached to the signals. Confidence scores are retained for diagnostics but are not used directly in the portfolio-weighting rule.
\end{itemize}

The Hawkish Agent is instructed to place greater weight on inflation pressure, tight financial conditions, and elevated real rates. The Dovish Agent is instructed to place greater weight on employment, industrial recovery, and the possibility that inflation pressure is temporary. Both agents observe the same macro evidence; the only difference is the interpretive prior through which that evidence is evaluated. All calls use temperature equal to zero. Deterministic decoding reduces sampling variability, and outputs are cached by date and agent type so that the backtest uses a fixed set of agent contracts. The full prompt templates used for the Hawkish, Dovish, and Debate Agents are reported in Appendix~\ref{app:prompts}.

\subsection{Debate Mechanism}

Let $C^{H}_{t}$ and $C^{D}_{t}$ denote the first-round contracts from the Hawkish and Dovish agents, respectively. Let $s^{H}_{i,t}$ and $s^{D}_{i,t}$ denote the corresponding ticker-level tilt signals for asset $i$. We measure the average absolute disagreement between the two agents as
\begin{equation}
\delta_t
=
\frac{1}{N}
\sum_{i=1}^{N}
\left|s^{H}_{i,t} - s^{D}_{i,t}\right|,
\label{eq:divergence}
\end{equation}
where $N=15$ is the number of ETFs in the investment universe.

If $\delta_t \leq \theta = 0.15$, the two contracts are merged without an additional deliberation round. If disagreement is larger, each agent reviews the other agent's contract and may revise its own ticker-level tilt signals, subject to a maximum of $R=2$ deliberation rounds. The final Debate Agent tilt signal is defined as the equal-weighted average of the revised Hawkish and Dovish tilt signals:
\begin{equation}
s^{\mathrm{Debate}}_{i,t}
=
\frac{1}{2}s^{H,\mathrm{rev}}_{i,t}
+
\frac{1}{2}s^{D,\mathrm{rev}}_{i,t}.
\label{eq:debate_tilt}
\end{equation}

This structure is intentionally simple. It allows the results to show whether the debate procedure creates additional value or whether the consensus mainly reflects averaging between two interpretive macro priors. The disagreement threshold $\theta$ and the maximum number of deliberation rounds $R$ are fixed ex ante and are not estimated from evaluation-period returns.

\subsection{Portfolio Weight Construction}
\label{sec:weights}

Portfolio weights are constructed using a common portfolio engine across all strategies. Let $S$ denote a signal-generating strategy, with
\[
S \in \{\mathrm{Rule}, \mathrm{Hawkish}, \mathrm{Dovish}, \mathrm{Debate}\}.
\]
The passive inverse-volatility benchmark is obtained by setting all macro tilt signals to zero. This common engine ensures that differences in performance come from the macro-to-tilt mapping rather than from differences in portfolio construction.

First, the inverse-volatility base weight uses a 26-week rolling volatility estimate:
\begin{equation}
w^{\mathrm{ivol}}_{i,t}
=
\frac{1/\widehat{\sigma}_{i,t}}
{\sum_{j=1}^{N} 1/\widehat{\sigma}_{j,t}},
\label{eq:ivol}
\end{equation}
where $\widehat{\sigma}_{i,t}$ is the rolling volatility estimate for asset $i$ and $N=15$ is the number of ETFs in the investment universe. Second, this inverse-volatility allocation is blended with equal weights:
\begin{equation}
w^{\mathrm{base}}_{i,t}
=
\alpha_{\mathrm{EW}} w^{\mathrm{ivol}}_{i,t}
+
(1-\alpha_{\mathrm{EW}})\frac{1}{N},
\qquad
\alpha_{\mathrm{EW}}=0.50.
\label{eq:blend}
\end{equation}
Third, strategy-specific macro tilts are applied multiplicatively to the base allocation:
\begin{equation}
\widetilde{w}^{S}_{i,t}
=
w^{\mathrm{base}}_{i,t}
\left(1+\kappa s^{S}_{i,t}\right),
\qquad
\kappa=0.25,
\label{eq:tilt}
\end{equation}
where $s^{S}_{i,t}\in[-2,+2]$ is the final ticker-level tilt signal produced by strategy $S$. Since $\kappa=0.25$, the tilt multiplier lies between $0.50$ and $1.50$, so macro signals adjust but do not dominate the base allocation. Then, the portfolio is subject to a risk-off cyclical cap. Let $\widehat{\rho}^{S}_{\mathrm{ro},t}$ denote the estimated risk-off probability for strategy $S$, and let $\mathcal{C}$ denote the pre-specified set of cyclical commodity ETFs. When
\[
\widehat{\rho}^{S}_{\mathrm{ro},t} > 0.65,
\]
the total portfolio weight assigned to cyclical assets is capped at
\begin{equation}
\sum_{i\in\mathcal{C}} w^{S}_{i,t}
\leq
\overline{w}_{\mathrm{cyc}},
\qquad
\overline{w}_{\mathrm{cyc}}=0.45.
\label{eq:cyclical_cap}
\end{equation}
The cap is applied by proportionally scaling down cyclical positions when the constraint binds, followed by renormalization of the full portfolio. Weights are renormalized and constrained by a maximum single-name weight and a maximum weekly turnover rule:
\begin{equation}
w^{S}_{i,t} \leq \overline{w}_{i},
\qquad
\overline{w}_{i}=0.30,
\label{eq:single_name_cap}
\end{equation}
and
\begin{equation}
\Delta^{S}_{t}
=
\frac{1}{2}
\sum_{i=1}^{N}
\left|w^{S}_{i,t}-w^{S}_{i,t-1}\right|
\leq
0.50.
\label{eq:turnover_cap}
\end{equation}
If the turnover constraint binds, the target portfolio is scaled back toward the previous week's weights until the constraint is satisfied. All weights formed at date $t$ are applied to returns over the following week, which imposes a one-week execution lag.

\subsection{Performance Measurement and Inference}

We evaluate each strategy using weekly portfolio returns. Let $r^S_t$ denote the weekly return of strategy $S$ at date $t$. We report annualised return, annualised volatility, Sharpe ratio, maximum drawdown, and hit rate. Annualised return and volatility are computed from weekly returns as
\begin{equation}
\mathrm{AnnRet}^S
=
52 \cdot \bar{r}^S,
\qquad
\mathrm{AnnVol}^S
=
\sqrt{52}\cdot \widehat{\sigma}(r^S_t),
\label{eq:annualization}
\end{equation}
where $\bar{r}^S$ is the sample mean weekly return and $\widehat{\sigma}(r^S_t)$ is the sample standard deviation of weekly returns.

The Sharpe ratio is defined as
\begin{equation}
\mathrm{Sharpe}^S
=
\frac{\mathrm{AnnRet}^S}{\mathrm{AnnVol}^S}.
\label{eq:sharpe}
\end{equation}
We do not subtract the risk-free rate in the baseline Sharpe calculation. This convention is used to maintain consistency with the strategy-level return comparison and is applied uniformly across all strategies. As a result, the reported Sharpe ratios should be interpreted as return-to-volatility ratios rather than risk-free-adjusted Sharpe ratios.

For a Sharpe difference between strategies $A$ and $B$,
\[
\Delta SR = \mathrm{Sharpe}^{A} - \mathrm{Sharpe}^{B},
\]
we use a paired stationary bootstrap with $B=5{,}000$ resamples and automatic block-length selection. The bootstrap is applied to the joint weekly return vector across strategies, which preserves the contemporaneous dependence between strategy returns as well as the time-series dependence induced by overlapping macro conditions and portfolio constraints. Bootstrap confidence intervals and $p$-values are then computed from the empirical distribution of the resampled Sharpe differences.

Transaction costs are evaluated by subtracting one-way trading costs from weekly portfolio returns. Let $c$ denote the one-way cost in basis points, with
\[
c \in \{0,5,10,20,30\}.
\]
The net return after transaction costs is
\begin{equation}
r^{S,c}_t
=
r^S_t
-
(c \times 10^{-4})\Delta^S_t,
\qquad
\Delta^S_t
=
\frac{1}{2}
\sum_{i=1}^{N}
\left|w^S_{i,t}-w^S_{i,t-1}\right|,
\label{eq:transaction_cost}
\end{equation}
where $\Delta^S_t$ is the one-way portfolio turnover implied by the weekly rebalancing from $w^S_{t-1}$ to $w^S_t$. This transaction-cost exercise is intended as a sensitivity analysis rather than a full implementation-cost model, since it does not include market impact, taxes, or operational frictions.


\section{Data}
\label{sec:data}

\subsection{Commodity-Related ETF Universe}

The investment universe contains 15 U.S.-listed commodity-related ETFs, covering precious metals, energy, agriculture, transition metals, broad commodity exposures, and one equity cash-flow proxy. The universe is therefore best interpreted as a commodity-related allocation universe rather than a pure commodity-only universe. Most assets provide direct commodity or commodity-subsector exposure. COWZ is retained because it was included in the pre-specified universe as an equity cash-flow proxy linked to real-asset and inflation-sensitive exposure; it is not interpreted as direct commodity exposure. Accordingly, all empirical results are framed as evidence on commodity-related ETF allocation rather than pure commodity investing.

\subsection{Macroeconomic Data}

The macro feature set contains seven FRED series, listed in Table~\ref{tab:macro}.
These variables are selected because they correspond to the main macro channels emphasized in the commodity literature: risk appetite, dollar strength, monetary policy, industrial demand, inflation expectations, real yields, and labor-market slack.
Daily series are used as observed at the weekly close.
Monthly series are lagged to reflect public release timing.

\subsection{LLM Configuration}
\label{sec:llmconfig}

All LLM calls use \texttt{gpt-4o-mini} through the OpenAI API with temperature set to zero. The zero-temperature setting is used to reduce sampling variability in the mapping from standardized macro evidence to portfolio tilts. To ensure that the backtest is based on a fixed set of model outputs, all generated contracts are cached by date and agent type. The prompts provide only standardized macro states and mechanically constructed summaries based on those same variables; they do not provide future returns, future macro realizations, unrestricted news text, or external tool access. Based on observed token usage and API prices at the time of implementation, the direct API cost is approximately \$0.002 per weekly observation. This figure excludes the operational burden of maintaining prompts, validating JSON outputs, and monitoring model behavior.

\subsection{Evaluation Window and Sub-Period Definition}

The evaluation sample contains 124 weekly rebalancing dates from October 2023 through February 2026. Observations before the evaluation window are used only to initialise rolling z-scores and volatility estimates. For descriptive sub-period analysis, we divide the sample into two calendar-based macro regimes. The \emph{Rates Peak} period covers the late-2023 environment, when U.S. policy rates and real yields were elevated and monetary tightening remained the dominant macro theme. The \emph{Soft Landing} period covers January 2024 through February 2026, when inflation moderated while economic activity remained resilient and macro signals became more mixed.

These sub-period labels are used only for ex post performance attribution and are not used by any strategy in forming portfolio weights. The division is useful because it separates a tightening-dominated environment from a later period in which inflation, growth, and rate signals provided less one-sided guidance.

\section{Empirical Results}
\label{sec:results}
\subsection{Full-Period Performance}

Table 3 reports the full-period performance of the five strategies. The inverse-volatility benchmark earns an annualized return of 6.68\% with a Sharpe ratio of 0.52, while the Rule Agent earns 7.11\% with a Sharpe ratio of 0.53. All three LLM-based strategies deliver higher full-period return-to-volatility ratios than the Rule Agent. The Hawkish Agent has the strongest full-period performance, with an annualized return of 7.74\% and a Sharpe ratio of 0.57. The Debate Agent follows closely at 7.69\% and 0.57, while the Dovish Agent earns 7.61\% with a Sharpe ratio of 0.56.

Figure 1 shows that these performance differences emerge mainly late in the sample. The cumulative-return paths remain close through late 2023 and much of 2024, while the relative performance gap becomes more visible during the soft-landing period's later recovery phase. The drawdown panel also shows that the LLM strategies do not avoid the main losses. Their advantage comes primarily from stronger participation in the recovery rather than from materially reducing downside risk.

\subsection{Incremental Value of LLM-Based Macro Interpretation}

Table 4 compares each LLM-based strategy with the Rule Agent. Relative to the Rule Agent, the LLM agents add 51–64 basis points of annualized return and 0.034–0.044 Sharpe points over the full sample. These effects are economically modest, but the controlled design makes them informative: the information set, risk constraints, and portfolio construction engine are held fixed, so any systematic differences are attributable to the mapping between standardized macro states and ticker-level tilt signals.

The Debate Agent's Sharpe ratio is only 0.001 above the arithmetic average of the Hawkish and Dovish Sharpe ratios. This provides little evidence of a separate deliberation premium in the full sample. Instead, the Debate Agent is better interpreted as a prior-aggregation mechanism that reduces dependence on either a single interpretive prior.

\subsection{Active Return Attribution}

Figure 2 decomposes the Debate Agent's active return relative to the Rule Agent. The largest positive contributions come from precious-metals exposures, especially SLV, GLD, and PALL, while several energy and agricultural positions, including BNO and CORN, detract from relative performance. Panel B shows that the Debate Agent's cumulative active return is negative for part of the sample but turns positive during the later recovery phase.

This pattern suggests that the Debate Agent's relative performance is driven more by cross-sectional allocation across commodity-related ETFs than by a simple increase in broad commodity exposure. This interpretation is consistent with the paper's focus on the macro-to-tilt mapping.

\subsection{Bootstrap Significance Tests}

Table 5 reports paired stationary block-bootstrap tests for pairwise Sharpe-ratio differences. The bootstrap resamples the joint weekly return vector across strategies, rather than each strategy separately, so that both time-series dependence and contemporaneous cross-strategy dependence are preserved. This is important because the weekly portfolio returns are generated from overlapping macro conditions, rolling volatility estimates, and turnover-constrained portfolio weights.

Hawkish versus Rule is positive and statistically different from zero at the 10\% level under the unadjusted bootstrap test ($p=0.067$), and Debate versus Rule also reaches the 10\% level ($p=0.089$). The Dovish comparison is positive but not statistically significant, and no cross-agent comparison is significant. Because the sample contains only 124 weekly observations and several related pairwise tests are reported, these results should be interpreted as directional evidence rather than conclusive inference. 

\subsection{Sub-Period Analysis}

Table 6 decomposes performance by macro regime. During the Rates Peak period, all signal-based strategies have negative Sharpe ratios, while the inverse-volatility benchmark has the least negative Sharpe ratio. This pattern is consistent with the view that macro tilts were less useful in the late-2023 environment, when tight monetary policy and elevated real yields dominated commodity-related returns.

During the Soft Landing period, the ranking reverses. The Hawkish and Debate Agents deliver the highest Sharpe ratios, followed by the Dovish Agent, the Rule Agent, and the inverse-volatility benchmark. The sub-period evidence therefore suggests that the LLM-based contribution is concentrated in the period when inflation, growth, and rate signals gave more mixed guidance.

\subsection{Transaction Cost Sensitivity}

Table 7 and Figure 3 report transaction-cost sensitivity. For each strategy, transaction costs are applied to the realized weekly one-way turnover, and Sharpe ratios are recomputed from the resulting net weekly returns. The LLM-based strategies remain above the inverse-volatility benchmark across one-way cost assumptions from 0 to 30 basis points, while the Rule Agent's small advantage over the passive benchmark is largely eliminated at low cost levels. Figure 3 presents the same pattern visually through the cost-sensitivity curves and heatmap.

This exercise should be interpreted as a net-performance sensitivity rather than as evidence that the LLM-based strategies trade less, because the paper does not separately report strategy-level turnover statistics. The result also does not establish full implementability, since market impact, taxes, bid--ask spread variation, and operational frictions are excluded. It shows only that the LLM-based advantage is not purely a zero-cost artifact under the realized turnover paths.

\subsection{Debate Mechanism Dynamics}

Figure 4 examines the behavior of the debate mechanism. The two single agents often disagree only modestly, as reflected in the low average divergence between their portfolio weights and regime assessments. Larger disagreements occur around macro turning points, especially when inflation and growth signals provide conflicting guidance.

This pattern helps explain both the stability and the limitations of the Debate Agent. The Debate Agent primarily serves as a conservative averaging device between the Hawkish and Dovish interpretations. It should therefore be interpreted as a prior-aggregation mechanism rather than as a separate information-acquisition channel or an independent signal source.

\subsection{Risk Profile Robustness}

Table 8 compares the risk-neutral and risk-averse parameterizations of the portfolio engine. The risk-averse profile imposes a tighter cyclical cap and a more conservative risk-off trigger, while the risk-neutral profile allows greater cyclical exposure. The LLM-based strategies maintain positive Sharpe differences relative to the Rule Agent under both profiles.

The ranking of the single agents changes across profiles. The Hawkish Agent performs best under the risk-neutral profile, while the Dovish Agent improves under the risk-averse profile. The Debate Agent is the least sensitive to the profile change. This stability is consistent with the interpretation that consensus reduces dependence on any single macro prior when the appropriate regime interpretation is uncertain.


\section{Discussion}
\label{sec:discussion}

\subsection{Does LLM-Based Macro Interpretation Add Value Over a Rule Layer?}

The evidence is positive but modest. LLM-based strategies produce higher Sharpe ratios than the Rule Agent across the main specifications, but the improvements are small and statistically fragile. The strongest comparisons reach the 10\% level in the paired stationary block bootstrap, not the 5\% level, and the confidence intervals include zero. Thus, the paper does not show that LLMs can reliably generate commodity trading profits in general.

The result is narrower: in one U.S. rate-cycle sample, with a fixed information set and a fixed implementation rule, LLM-based macro interpretation delivers modestly higher return-to-volatility performance than a transparent sign-mapping benchmark.

\subsection{Interpretive Prior and Regime Alignment}

The Hawkish--Dovish comparison shows that an LLM-based interpretive prior is most useful when it is aligned with the prevailing macro regime. The sample begins in an environment of elevated real rates and tight monetary policy, which is more naturally aligned with the Hawkish prior. As conditions shift toward a soft landing, the Dovish interpretation becomes more competitive, although it can also generate excessive cyclical exposure under the less restrictive risk-neutral parameterisation.

The Debate Agent partly mitigates this prior-selection problem by averaging the Hawkish and Dovish views. However, the results do not show that deliberation creates a separate signal beyond the two underlying priors. A natural extension is dynamic prior weighting, in which the Hawkish or Dovish Agent receives greater weight as estimated regime probabilities change.

\subsection{Limitations}

Five limitations are central. First, the backtest covers only one U.S. rate cycle, and the soft-landing period drives much of the full-sample advantage. Second, the macro data are release-aware but not fully vintage; an ALFRED-based implementation would better reconstruct the information set available to investors in real time. Third, the asset universe is commodity-related rather than purely commodity-only because it includes one equity cash-flow proxy. Future robustness work should re-estimate the framework after excluding this proxy.

Fourth, the prompt protocol does not yet include a masked-date robustness test, which would further reduce concerns that the pretrained model uses calendar-specific background knowledge rather than only the supplied macro state. Fifth, the portfolio construction parameters are fixed before the evaluation period but are still chosen by the researchers, and the reported bootstrap p-values are not formally adjusted for multiple testing.

Longer out-of-sample tests, fully vintage macro data, masked-date prompts, pure-commodity universe checks, and systematic parameter sensitivity analysis would help determine whether the observed advantage reflects a persistent interpretation effect or a favorable sample period.


\section{Conclusion}
\label{sec:conclusion}

This paper examines whether LLMs add value to commodity-related ETF portfolio construction when their role is limited to macro interpretation. All strategies observe the same structured macro data and use the same portfolio engine; only the mapping from macro states to ticker-level tilt signals differs. This controlled design allows the paper to evaluate LLM-based macro interpretation as a constrained interpretation layer rather than as an unrestricted trading system.

The results are cautiously affirmative. The Hawkish, Dovish, and Debate Agents all deliver higher full-sample return-to-volatility ratios than the deterministic Rule Agent, with the strongest evidence for the Hawkish and Debate Agents. The gains are concentrated in the soft-landing period and remain visible under moderate one-way transaction-cost assumptions. However, the statistical strength is modest, and the strongest comparisons do not survive conservative multiple-testing adjustment.

The Debate Agent is useful mainly as a prior-averaging device. It does not outperform the strongest single agent, so the data do not support a separate deliberation premium in this sample. Its value is instead stability: it reduces reliance on any one interpretive macro prior when the appropriate regime interpretation is uncertain.

The broader implication is that LLMs are more credible as constrained interpretation tools than as standalone trading engines. In this setting, LLM-based macro interpretation adds a small improvement over a transparent rule-based benchmark. Whether that improvement is persistent enough to justify the added operational complexity requires longer samples, fully vintage macro data, masked-date robustness tests, pure-commodity universe checks, and further out-of-sample evaluation.


\section*{Funding details}

The authors received no financial support for the research, authorship, and/or publication of this article.

\section*{Disclosure statement}

The authors report there are no competing interests to declare. The views expressed in this paper are solely those of the author and do not represent the views, positions, or policies of Citigroup Inc. or any of its affiliates.

\section*{Author Contributions statement}

Kerui Geng contributed to the conception and design of the study, the empirical methodology, the analysis and interpretation of the data, and the drafting and revision of the manuscript. Yiqing Wang contributed to literature review, the research design, the LLM-agent framework, interpretation of results, and manuscript revision. Dehao Dai contributed to data processing, implementation support, empirical validation, interpretation of the results, and critical revision of the manuscript for intellectual content. Ding Ma contributed to portfolio construction, robustness analysis, interpretation of empirical findings, and manuscript revision. All authors reviewed and approved the final version of the manuscript and agree to be accountable for all aspects of the work.

\section*{Data availability statement}

The raw market data are obtained from Yahoo Finance, and the macroeconomic variables are obtained from FRED. The processed weekly panel, generated LLM contracts, and replication materials that support the findings of this study are available from the corresponding author upon reasonable request, subject to any applicable data-use restrictions.

\section*{Generative AI use statement}

For manuscript preparation, the authors used ChatGPT (GPT-5.5 Thinking, OpenAI) only for writing polishing and editorial assistance, including grammar, clarity, concision, and formatting. This use is separate from the LLM agents analyzed as part of the research design, which are described in Sections~\ref{sec:methodology} and~\ref{sec:data}. The authors reviewed and edited all AI-assisted text and take full responsibility for the final content of the manuscript.


\bibliographystyle{apalike}
\bibliography{main}
\clearpage
\setcounter{table}{0}

\begin{table}[!ht]
\caption{Commodity-Related ETF Universe}
\label{tab:etfs}
\centering\small
\begin{threeparttable}
\begin{tabular}{llll}
\toprule
Ticker & Name & Sub-class & Exchange \\
\midrule
GLD  & SPDR Gold Shares                          & Precious Metals    & NYSE Arca \\
SLV  & iShares Silver Trust                      & Precious Metals    & NYSE Arca \\
PALL & Aberdeen Physical Palladium Shares ETF    & Precious Metals    & NYSE Arca \\
TMET & iShares Transition-Enabling Metals ETF    & Transition Metals  & NASDAQ    \\
USO  & United States Oil Fund                    & Energy             & NYSE Arca \\
BNO  & United States Brent Oil Fund              & Energy             & NYSE Arca \\
DBO  & Invesco DB Oil Fund                       & Energy             & NYSE Arca \\
GSG  & iShares S\&P GSCI Commodity Index ETF     & Broad Commodity    & NYSE Arca \\
PDBC & Invesco Optimum Yield Diversified ETF     & Broad Commodity    & NASDAQ    \\
FTGC & First Trust Global Tactical Commodity ETF & Broad Commodity    & NASDAQ    \\
BCI  & Aberdeen Bloomberg All Commodity ETF      & Broad Commodity    & NYSE Arca \\
COWZ & Pacer US Cash Cows 100 ETF               & Equity Cash-Flow Proxy & CBOE BZX  \\
CORN & Teucrium Corn Fund                        & Agriculture        & NYSE Arca \\
WEAT & Teucrium Wheat Fund                       & Agriculture        & NYSE Arca \\
SOYB & Teucrium Soybean Fund                     & Agriculture        & NYSE Arca \\
\bottomrule
\end{tabular}
\begin{tablenotes}\small
\item Price data are sourced from Yahoo Finance. The price-history window begins in October 2022 to initialise rolling z-score and volatility estimates; the performance evaluation begins in October 2023 and ends in February 2026. COWZ is included as an equity cash-flow proxy in the original pre-specified universe and should not be interpreted as direct commodity exposure.
\end{tablenotes}
\end{threeparttable}
\end{table}

\begin{table}[!ht]
\caption{Macroeconomic Feature Set}
\label{tab:macro}
\centering\small
\begin{threeparttable}
\begin{tabular}{llL{5.5cm}c}
\toprule
Feature & FRED Series & Description & Release Lag (weeks) \\
\midrule
$z^{\text{vix}}$    & VIXCLS    & CBOE Volatility Index                                & 0 \\
$z^{\text{usd}}$    & DTWEXBGS  & Broad nominal U.S.\ dollar index                    & 0 \\
$z^{\text{ff}}$     & FEDFUNDS  & Effective federal funds rate                        & 0 \\
$z^{\text{indpro}}$ & INDPRO    & Industrial Production Index (seasonally adjusted)   & 2 \\
$z^{\text{bkevn}}$  & T10YIE    & 10-year breakeven inflation rate                    & 0 \\
$z^{\text{ryr}}$    & DFII10    & 10-year TIPS real yield                             & 0 \\
$z^{\text{unrate}}$ & UNRATE    & Civilian unemployment rate (seasonally adjusted)    & 2 \\
\bottomrule
\end{tabular}
\begin{tablenotes}\small
\item Monthly indicators are aligned to the latest public release available as of each weekly rebalancing date. Release-lag corrections are applied prior to z-score computation.
\end{tablenotes}
\end{threeparttable}
\end{table}

\begin{table}[ht]
\caption{Full-Period Performance Summary}
\label{tab:fullperiod}
\centering\small
\begin{threeparttable}
\begin{tabular}{lrrrrr}
\toprule
Strategy & Ann.\ Return & Ann.\ Vol. & Sharpe & Max DD & Hit Rate \\
\midrule
Rule Agent (z-score)  & 7.11\% & 13.50\% & 0.53 & $-$9.15\% & 55.65\% \\
Hawkish Agent         & 7.74\% & 13.57\% & 0.57 & $-$9.37\% & 55.65\% \\
Dovish Agent          & 7.61\% & 13.57\% & 0.56 & $-$9.48\% & 54.84\% \\
Debate Agent          & 7.69\% & 13.57\% & 0.57 & $-$9.40\% & 54.84\% \\
Inverse Volatility$^*$& 6.68\% & 12.81\% & 0.52 & $-$9.54\% & 56.45\% \\
\bottomrule
\end{tabular}
\begin{tablenotes}\small
\item[$^*$] Inverse volatility is the passive risk-only benchmark.
\item Sharpe ratios are reported without subtracting the risk-free rate, consistent with the commodity roll-yield convention. Sample: 124 weekly observations, October 2023--February 2026.
\end{tablenotes}
\end{threeparttable}
\end{table}

\begin{table}[ht]
\caption{Incremental Value of LLM Strategies Relative to the Rule Agent}
\label{tab:incremental}
\centering\small
\begin{threeparttable}
\begin{tabular}{lrr}
\toprule
Strategy & $\Delta$ Ann.\ Return & $\Delta$ Sharpe \\
\midrule
Hawkish Agent                       & $+0.64\%$ & $+0.044$ \\
Dovish Agent                        & $+0.51\%$ & $+0.034$ \\
Debate Agent                        & $+0.58\%$ & $+0.040$ \\
Debate vs.\ avg(Hawkish, Dovish)    & ---       & $+0.001$ \\
\bottomrule
\end{tabular}
\begin{tablenotes}\small
\item Incremental annualised return and Sharpe ratio relative to the Rule Agent, over the full 124-week sample.
The last row compares the Debate Agent against the arithmetic average of the two single-agent Sharpe ratios.
\end{tablenotes}
\end{threeparttable}
\end{table}

\begin{table}[ht]
\caption{Block-Bootstrap Sharpe Difference Tests}
\label{tab:bootstrap}
\centering\small
\begin{threeparttable}
\begin{tabular}{lrrrr}
\toprule
Comparison & $\Delta$Sharpe & $p$-value & 95\% CI & Sig. \\
\midrule
Hawkish vs.\ Rule          & $+0.0415$ & 0.067 & $[-0.014,\; +0.092]$ & $*$   \\
Dovish vs.\ Rule           & $+0.0323$ & 0.134 & $[-0.025,\; +0.090]$ & (ns)  \\
Debate vs.\ Rule           & $+0.0379$ & 0.089 & $[-0.017,\; +0.092]$ & $*$   \\
Debate vs.\ Hawkish        & $-0.0037$ & 0.769 & $[-0.013,\; +0.007]$ & (ns)  \\
Debate vs.\ Dovish         & $+0.0056$ & 0.163 & $[-0.006,\; +0.016]$ & (ns)  \\
Any LLM vs.\ Inv.\ Vol.\   & $+0.0454$ & 0.262 & $[-0.101,\; +0.180]$ & (ns)  \\
Rule vs.\ Inv.\ Vol.\      & $+0.0075$ & 0.452 & $[-0.123,\; +0.130]$ & (ns)  \\
Hawkish vs.\ Dovish        & $+0.0092$ & 0.183 & $[-0.012,\; +0.028]$ & (ns)  \\
\bottomrule
\end{tabular}
\begin{tablenotes}\small
\item Stationary bootstrap with $B = 5{,}000$ resamples and automatic block length selection.
$*$ denotes significance at the 10\% level; (ns) = not significant.
$p$-values are unadjusted for multiple comparisons; see the text for a Bonferroni discussion.
\end{tablenotes}
\end{threeparttable}
\end{table}

\begin{table}[ht]
\caption{Sub-Period Sharpe Ratios}
\label{tab:subperiod}
\centering\small
\begin{threeparttable}
\begin{tabular}{lrrrrr}
\toprule
Period & Rule & Hawkish & Dovish & Debate & Inv.\ Vol. \\
\midrule
Rates Peak (2023)     & $-1.46$ & $-1.33$ & $-1.32$ & $-1.32$ & $-1.30$ \\
Soft Landing (2024--25)&  0.84  &   0.86  &   0.85  &   0.86  &   0.80  \\
Full Period            &  0.53  &   0.57  &   0.56  &   0.57  &   0.52  \\
\bottomrule
\end{tabular}
\begin{tablenotes}\small
\item Sub-period boundaries follow the U.S.\ policy rate cycle.
Rates Peak = 2023; Soft Landing = 2024--2025.
\end{tablenotes}
\end{threeparttable}
\end{table}

\begin{table}[ht]
\caption{Transaction Cost Sensitivity}
\label{tab:txcost}
\centering\small
\begin{threeparttable}
\begin{tabular}{lrrrrr}
\toprule
Strategy & 0 bps & 5 bps & 10 bps & 20 bps & 30 bps \\
\midrule
Rule Agent (z-score) & 0.526 & 0.520 & 0.514 & 0.503 & 0.491 \\
Hawkish Agent        & 0.571 & 0.567 & 0.562 & 0.554 & 0.546 \\
Dovish Agent         & 0.561 & 0.556 & 0.550 & 0.540 & 0.530 \\
Debate Agent         & 0.567 & 0.562 & 0.558 & 0.549 & 0.539 \\
Inverse Volatility   & 0.521 & 0.514 & 0.508 & 0.494 & 0.481 \\
\bottomrule
\end{tabular}
\begin{tablenotes}\small
\item One-way cost applied to weekly L1 turnover at each rebalancing date.
Hawkish and Debate Agents remain above the inverse-volatility benchmark across all cost scenarios.
The Rule Agent breaks even against passive at approximately 5 bps.
\end{tablenotes}
\end{threeparttable}
\end{table}

\begin{table}[ht]
\caption{Risk Profile Robustness: Full-Period Performance}
\label{tab:riskprofile}
\centering\small
\begin{threeparttable}
\begin{tabular}{lrrrrrrr}
\toprule
 & \multicolumn{2}{c}{Risk-Neutral} & \multicolumn{2}{c}{Risk-Averse} & \multicolumn{2}{c}{Difference (RA$-$RN)} & \\
\cmidrule(lr){2-3}\cmidrule(lr){4-5}\cmidrule(lr){6-7}
Strategy & Return & Sharpe & Return & Sharpe & $\Delta$Return & $\Delta$Sharpe & Consistent? \\
\midrule
Rule Agent      & 7.11\% & 0.53 & 7.11\% & 0.53 & $0.00\%$ & $+0.000$ & --- \\
Hawkish Agent   & 7.74\% & 0.57 & 7.60\% & 0.56 & $-0.14\%$ & $-0.010$ & Yes \\
Dovish Agent    & 7.61\% & 0.56 & 7.76\% & 0.57 & $+0.15\%$ & $+0.010$ & Yes \\
Debate Agent    & 7.69\% & 0.57 & 7.69\% & 0.57 & $0.00\%$  & $-0.000$ & Yes \\
Inverse Vol.    & 6.68\% & 0.52 & 6.68\% & 0.52 & $0.00\%$  & $+0.000$ & --- \\
\midrule
\multicolumn{8}{l}{\textit{LLM $\Delta$Sharpe vs.\ Rule Agent by profile}} \\
Hawkish Agent   & \multicolumn{2}{c}{$+0.044$} & \multicolumn{2}{c}{$+0.034$} & \multicolumn{3}{l}{consistent} \\
Dovish Agent    & \multicolumn{2}{c}{$+0.034$} & \multicolumn{2}{c}{$+0.044$} & \multicolumn{3}{l}{consistent} \\
Debate Agent    & \multicolumn{2}{c}{$+0.040$} & \multicolumn{2}{c}{$+0.040$} & \multicolumn{3}{l}{consistent} \\
\bottomrule
\end{tabular}
\begin{tablenotes}\small
\item Risk-neutral profile: cyclical cap $\in [0.25, 0.45]$, triggered at VIX z-score $> 1.5$.
Risk-averse profile: cyclical cap $\in [0.10, 0.25]$, triggered at any risk-off signal.
``Consistent'' denotes positive $\Delta$Sharpe vs.\ Rule Agent under both profiles.
\end{tablenotes}
\end{threeparttable}
\end{table}

\clearpage
\setcounter{figure}{0}

\begin{figure}[p]
\centering
\includegraphics[width=0.98\textwidth]{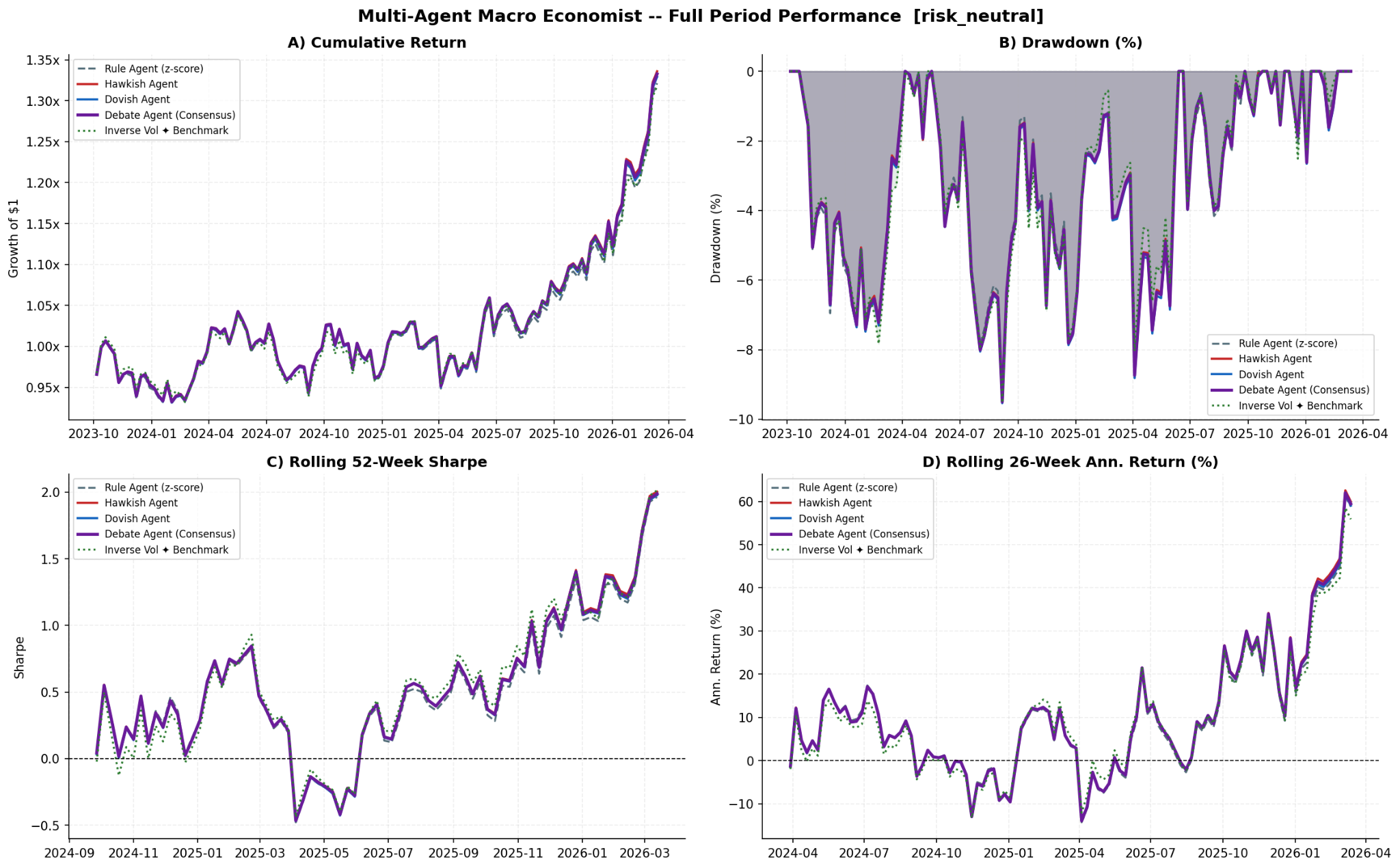}
\caption{Full-Period Strategy Performance}
\label{fig:performance_dashboard}
{\small \textit{Note:} The figure reports cumulative return, drawdown, rolling 52-week Sharpe ratio, and rolling 26-week annualised return for the Rule Agent, the three LLM agents, and the inverse-volatility benchmark. All portfolios are evaluated under the risk-neutral parameterisation. Sample: October 2023--February 2026.}
\end{figure}

\begin{figure}[p]
\centering
\includegraphics[width=0.98\textwidth]{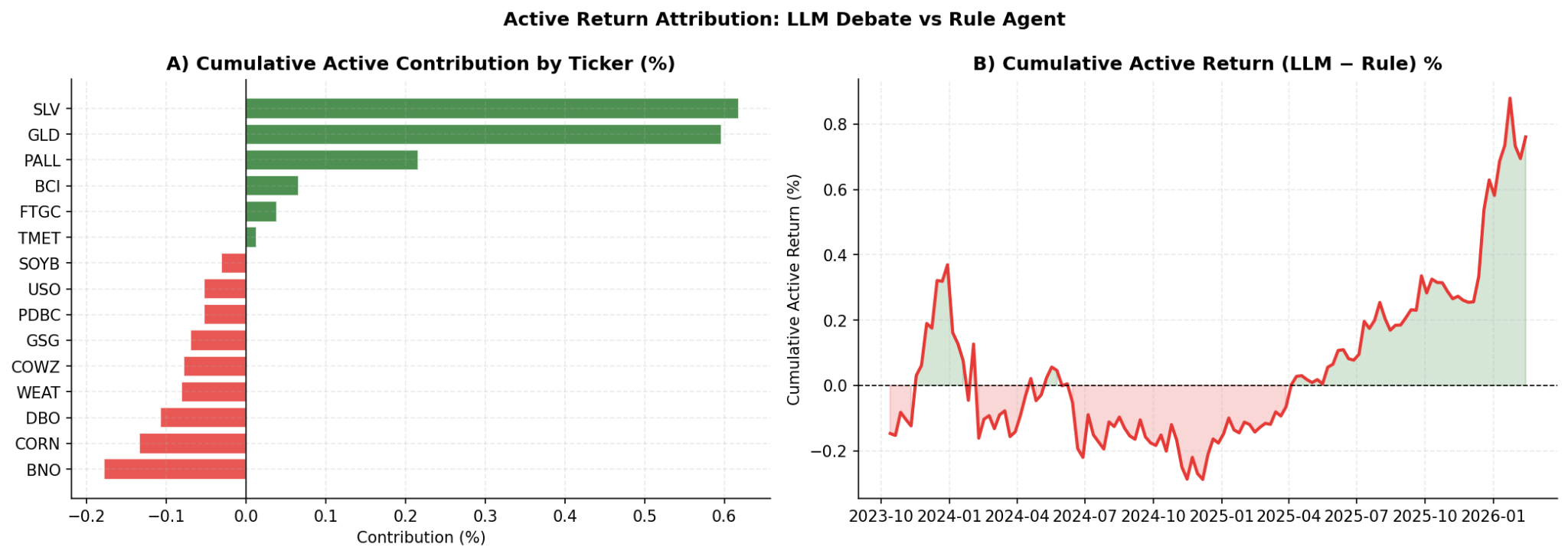}
\caption{Active Return Attribution: Debate Agent Relative to the Rule Agent}
\label{fig:active_attribution}
{\small \textit{Note:} Panel A decomposes the cumulative active return of the Debate Agent relative to the Rule Agent by ETF ticker. Panel B plots cumulative active return over time. Positive bars indicate tickers that contributed positively to the Debate Agent's active return; negative bars indicate detractors. Sample: October 2023--February 2026.}
\end{figure}

\begin{figure}[!ht]
\centering
\includegraphics[width=0.8\textwidth]{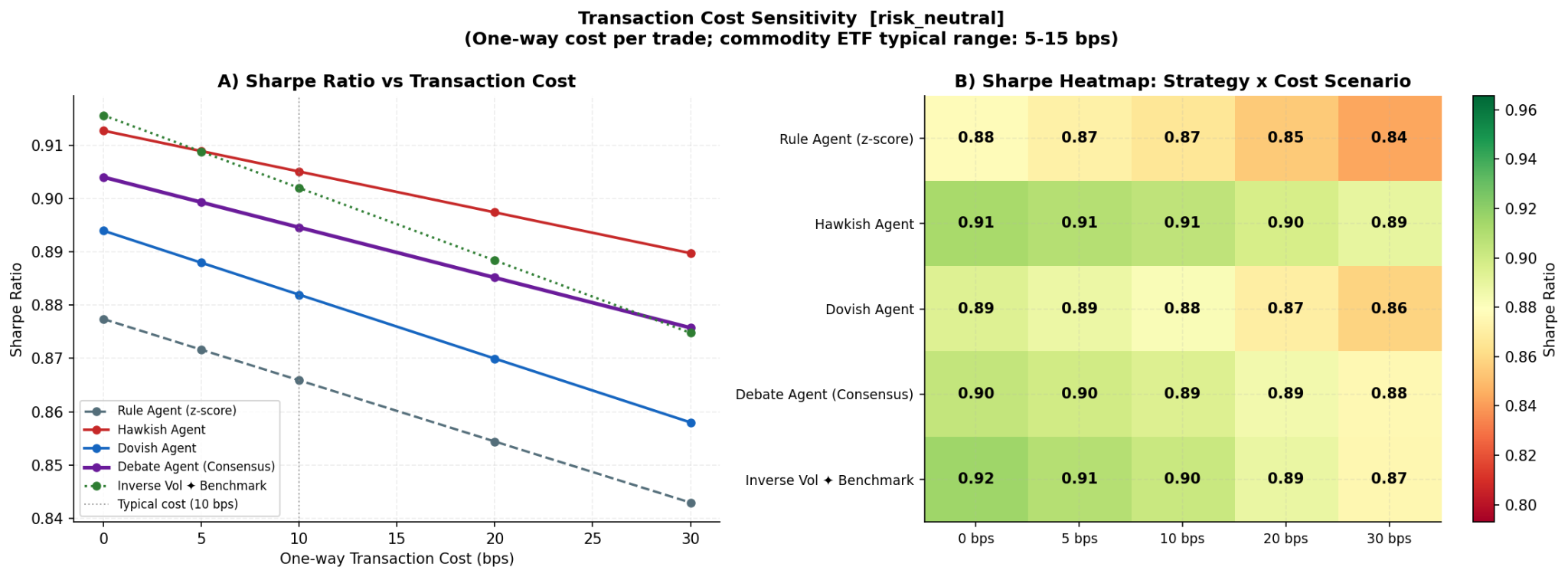}
\caption{Transaction Cost Sensitivity}
\label{fig:cost_sensitivity}
{\small \textit{Note:} Panel A plots Sharpe ratios as a function of one-way trading costs from 0 to 30 basis points. The vertical dotted line marks a representative 10 basis-point cost. Panel B reports the same information as a strategy-by-cost heatmap. All strategies are evaluated under the risk-neutral parameterisation.}
\end{figure}

\begin{figure}[!ht]
\centering
\includegraphics[width=0.8\textwidth]{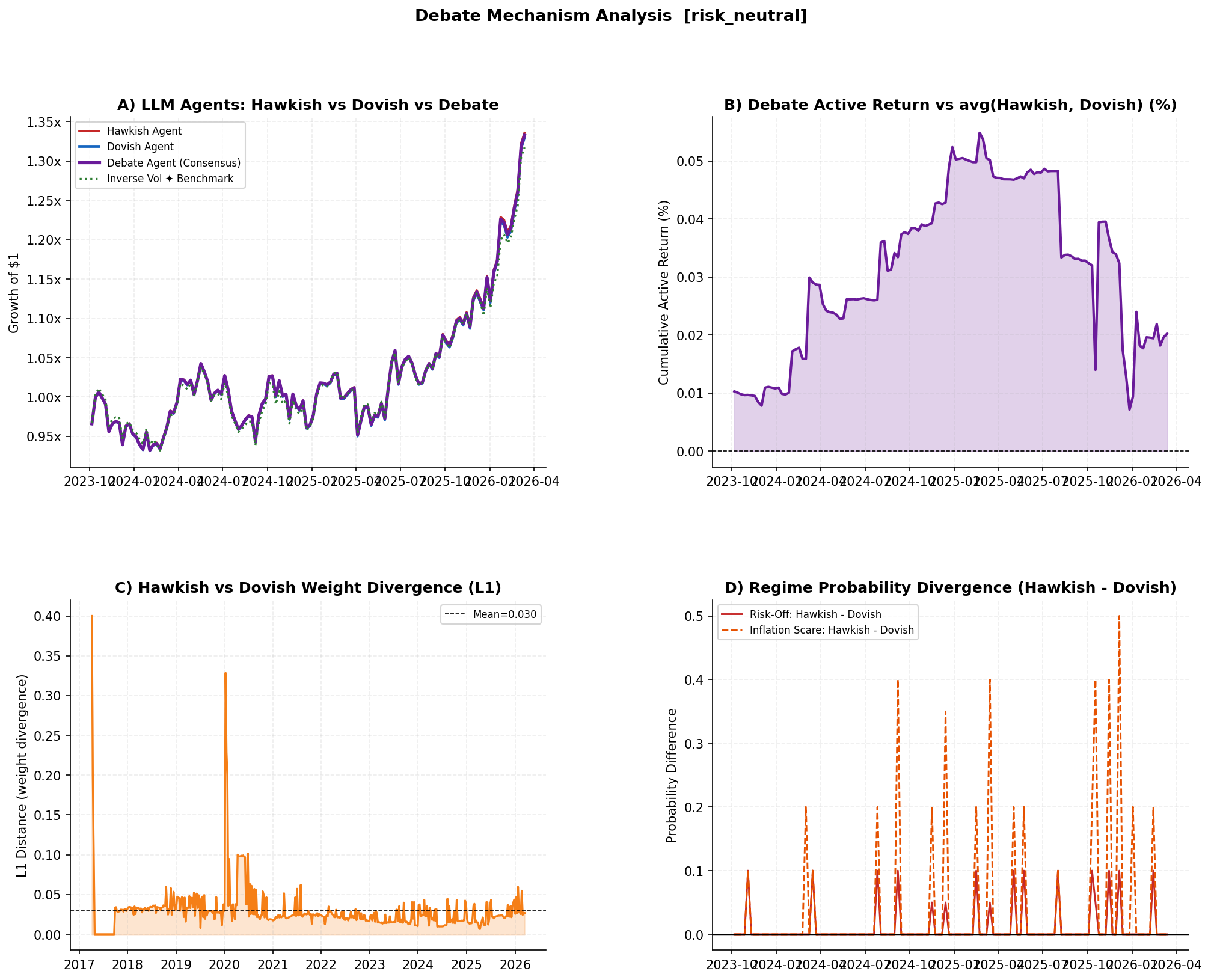}
\caption{Debate Mechanism Analysis}
\label{fig:debate}
{\small \textit{Note:} Panel A plots cumulative returns of the Hawkish, Dovish, and Debate Agents against the inverse-volatility benchmark. Panel B plots the cumulative active return of the Debate Agent relative to the simple average of Hawkish and Dovish returns. Panel C plots weekly L1 weight divergence between the Hawkish and Dovish agents, with the sample mean indicated by a dashed line. Panel D plots differences in estimated risk-off and inflation-scare regime probabilities between the Hawkish and Dovish agents.}
\end{figure}

\clearpage
\appendix
\renewcommand{\thesection}{\Alph{section}}
\renewcommand{\thetable}{\Alph{section}\arabic{table}}
\setcounter{table}{0}

\section{Rule Agent Z-Score Loading Matrix}
\label{app:loadings}

Table~\ref{tab:loadings} reports the directional loadings $\beta_{ik}$ used in equation~\eqref{eq:rule_raw_tilt}.
A positive (negative) sign indicates that an elevated z-score for the corresponding macro variable increases (decreases) the asset's tilt signal.
All loadings are derived from macro-finance theory; no loading is estimated from the return data used in the performance evaluation.

\begin{table}[ht]
\caption{Rule Agent Z-Score Loading Matrix ($\beta_{ik}$)}
\label{tab:loadings}
\centering\small
\begin{threeparttable}
\begin{tabular}{lccccccc}
\toprule
Asset & VIX & USD & FF & INDPRO & BKEVN & RYR & UNRATE \\
\midrule
GLD  & $+$ & $-$ & $-$ & $0$ & $+$ & $-$ & $+$ \\
SLV  & $+$ & $-$ & $-$ & $+$ & $+$ & $-$ & $0$ \\
PALL & $0$ & $-$ & $0$ & $+$ & $0$ & $0$ & $0$ \\
TMET & $+$ & $-$ & $-$ & $+$ & $+$ & $-$ & $0$ \\
USO  & $-$ & $-$ & $0$ & $+$ & $+$ & $0$ & $-$ \\
BNO  & $-$ & $-$ & $0$ & $+$ & $+$ & $0$ & $-$ \\
DBO  & $-$ & $-$ & $0$ & $+$ & $+$ & $0$ & $-$ \\
GSG  & $-$ & $-$ & $0$ & $+$ & $+$ & $-$ & $0$ \\
PDBC & $-$ & $-$ & $0$ & $+$ & $+$ & $-$ & $0$ \\
FTGC & $-$ & $-$ & $0$ & $+$ & $+$ & $-$ & $0$ \\
BCI  & $-$ & $-$ & $0$ & $+$ & $+$ & $-$ & $0$ \\
COWZ & $-$ & $0$ & $-$ & $+$ & $0$ & $-$ & $-$ \\
CORN & $0$ & $-$ & $0$ & $0$ & $+$ & $0$ & $0$ \\
WEAT & $0$ & $-$ & $0$ & $0$ & $+$ & $0$ & $0$ \\
SOYB & $0$ & $-$ & $0$ & $0$ & $+$ & $0$ & $0$ \\
\bottomrule
\end{tabular}
\begin{tablenotes}\small
\item `$+$', `$-$', and `$0$' denote positive, negative, and zero directional effects, respectively.
\end{tablenotes}
\end{threeparttable}
\end{table}

\section{Portfolio Construction Parameters}
\label{app:params}

\begin{table}[H]
\caption{Portfolio Construction Parameter Summary}
\label{tab:params}
\centering
\begin{threeparttable}
\begin{tabular}{llr}
\toprule
Parameter & Description & Value \\
\midrule
$\kappa$             & Macro tilt scaling factor              & 0.25 \\
$\alpha_{\text{EW}}$ & Equal-weight blend parameter           & 0.50 \\
$\bar{w}_i$          & Maximum weight per asset               & 0.30 \\
$\Delta_{\max}$      & Maximum L1 weekly turnover             & 0.50 \\
$\hat\rho^{\text{ro}}$ trigger & Risk-off probability trigger  & 0.65 \\
$\bar{w}_{\text{cyc}}$ & Cyclical weight cap in risk-off      & 0.45 \\
$\theta$             & Debate consensus threshold             & 0.15 \\
$R$                  & Maximum debate rounds                  & 2    \\
$W_z$                & Rolling z-score window (weeks)         & 156  \\
$W_\sigma$           & Rolling volatility window (weeks)      & 26   \\
LLM model            & OpenAI model identifier                & \texttt{gpt-4o-mini} \\
Temperature          & LLM decoding temperature               & 0.00 \\
\bottomrule
\end{tabular}
\begin{tablenotes}\small
\item All parameter values are fixed prior to the evaluation window.
No parameter is estimated from the return data used in the performance evaluation.
\end{tablenotes}
\end{threeparttable}
\end{table}

\section{Rule Agent: Full Specification}
\label{app:ruleagent}

\subsection{Conflict Detection}

Prior to computing asset-level tilt signals, the Rule Agent tests for internally contradictory macro evidence:
\begin{equation*}
\text{conflict}_t = \mathbf{1}\bigl[z^{\text{vix}}_t > 1 \,\wedge\, z^{\text{indpro}}_t > 1\bigr] \;\vee\; \mathbf{1}\bigl[z^{\text{bkevn}}_t > 1 \,\wedge\, z^{\text{ryr}}_t < -1\bigr].
\end{equation*}
The first clause flags simultaneously elevated fear and strong industrial production; the second flags elevated nominal inflation expectations alongside negative real yields.
When a conflict is detected, all tilt magnitudes are attenuated by $\gamma = 0.5$ to reflect regime uncertainty.

\subsection{Risk-Off Composite Score}

A portfolio-level risk-off indicator is computed as:
\begin{equation*}
\text{RiskOff}_t = 0.40\cdot\mathbf{1}[z^{\text{vix}}_t \geq 1.5] + 0.25\cdot\mathbf{1}[z^{\text{ff}}_t \geq 1.0] + 0.20\cdot\mathbf{1}[z^{\text{ryr}}_t \geq 1.0] + 0.15\cdot\mathbf{1}[z^{\text{usd}}_t \geq 0.5].
\end{equation*}
When $\text{RiskOff}_t \geq 0.65$, positive tilts for cyclical assets are suppressed to zero and the cyclical weight cap $\bar{w}_{\text{cyc}} = 0.45$ is enforced.

\subsection{Regime Probability Assignment}
The Rule Agent produces a deterministic regime probability vector over five regimes according to:
\begin{align}
\hat\rho^{\text{ro}}_t  &\propto \max(0,z^{\text{vix}}_t) + \max(0,z^{\text{ryr}}_t) + \max(0,z^{\text{usd}}_t), \notag\\
\hat\rho^{\text{is}}_t  &\propto \max(0,z^{\text{bkevn}}_t) + \max(0,z^{\text{ff}}_t), \notag\\
\hat\rho^{\text{sl}}_t  &\propto \max(0,z^{\text{indpro}}_t) + \max(0,-z^{\text{vix}}_t) + \max(0,-z^{\text{ff}}_t), \notag\\
\hat\rho^{\text{stag}}_t &\propto \max(0,z^{\text{bkevn}}_t)\cdot\max(0,-z^{\text{indpro}}_t). \notag
\end{align}
All values are clipped to $[0,1]$ and renormalized to unit sum; the risk-on probability is the residual.

\subsection{ Worked Example}

At the 2023 rates peak, a representative macro state has $z^{\text{vix}} = +1.8$, $z^{\text{ryr}} = +2.1$, $z^{\text{ff}} = +2.4$, $z^{\text{indpro}} = -0.6$, $z^{\text{bkevn}} = +0.3$.
The risk-off score evaluates to $0.40 + 0.25 + 0.20 + 0 = 0.85$, triggering the cyclical cap.
For GLD: $\tilde{s} = (+1)(1.8) + (-1)(2.1) + (-1)(2.4) = -2.7$, mapping to $b = -2$; since GLD is a defensive metal, the risk-off cap does not apply, and the signal survives---correctly reflecting the elevated opportunity cost of holding gold at peak real yields.
For USO: $\tilde{s} = (-1)(1.8) + (0)(0) + (-1)(0.6) = -2.4$, also mapping to $b = -2$; the risk-off cap further confirms the strong underweight.
The Rule Agent thus mechanically identifies the demand-destruction regime of 2023, consistent with the deeply negative Rates Peak Sharpe ratios reported in Table~\ref{tab:subperiod}.

\section{LLM Prompt Templates}
\label{app:prompts}

This appendix reports the prompt structure used for the LLM-based agents. The
weekly numerical inputs are filled in mechanically from the standardized macro
feature vector described in Section~3.2. The agents do not receive future returns,
realized backtest performance, benchmark results, unrestricted news text, or external
tool access. The Hawkish and Dovish Agents receive the same input block and differ
only in the macroeconomic prior stated in the role instruction. The Debate Agent
receives the two first-round contracts and forms a consensus under the same
information restrictions.

\subsection{Common Prompt Block}

\begin{verbatim}
You are a macroeconomist and cross-asset strategist. Your task is to
analyze the current macroeconomic state using only the provided macro
evidence table and macro feature summary. You must identify the most
relevant macro regime(s) and produce auditable portfolio policy signals
for the portfolio construction layer.
Hard rules:
1. Use only the provided numbers. Do not use external facts, news,
   invented data, or information not shown in the input.
2. Explicitly refer to the evidence table entries when explaining the
   macro interpretation. Relevant fields include driver, direction,
   z-score, and historical 95 percent band.
3. Do not use, infer, or request ETF returns, backtest performance,
   benchmark results, or future information.
4. Output must be valid JSON following the required schema.
\end{verbatim}

\subsection{Agent-Specific Role Instructions}

\begin{verbatim}
Hawkish Agent:
You place greater weight on inflation pressure, tight financial
conditions, elevated real rates, and U.S. dollar strength. When macro
evidence is mixed, interpret the state through a price-stability and
monetary-tightening prior.

Dovish Agent:
You place greater weight on employment conditions, industrial recovery,
easing financial conditions, and the possibility that inflation pressure
is temporary. When macro evidence is mixed, interpret the state through
a growth-supportive and recovery-oriented prior.

Debate Agent:
You review the first-round Hawkish and Dovish contracts and produce a
single consensus contract. Resolve disagreements by referring back to
the same macro evidence table. Do not use external information,
future information, ETF returns, or backtest performance.
\end{verbatim}

\subsection{Validation}

The JSON output is validated before entering the portfolio construction layer.
Regime probabilities are restricted to $[0,1]$, policy parameters are clipped to
their pre-specified bounds, and category adjustments are restricted to the asset
categories reported in Table~1. Textual rationales and evidence citations are retained
for auditability but are not directly used in portfolio weighting.

\end{document}